\documentclass[aps,prd,amsmath,amssymb,preprintnumbers,nofootinbib]{revtex4}
\pdfoutput=1
\usepackage{amsthm}
\usepackage{graphicx}
\usepackage{color}
\newcommand{\beq}{\begin{eqnarray}}
\newcommand{\eeq}{\end{eqnarray}}

\newcommand{\bmp}{\noindent\begin{minipage}{16cm}}
\newcommand{\emp}{\end{minipage}\vskip 7mm} 

\usepackage{dcolumn}
\usepackage{bm}
\usepackage{bbm}
\usepackage{subfigure}
\usepackage{pxfonts}

\theoremstyle{definition}

\theoremstyle{plain}

\usepackage{epsfig}

\usepackage[ margin=5pt, font=normalsize,labelfont=bf,justification=raggedright]{caption}
\usepackage{youngtab}
\usepackage{slashed}

\definecolor{rossoCP3}{cmyk}{0,.88,.77,.40}

\def\lsim{\mathrel{\rlap{\lower4pt\hbox{\hskip1pt$\sim$}}
    \raise1pt\hbox{$<$}}}                
\def\gsim{\mathrel{\rlap{\lower4pt\hbox{\hskip1pt$\sim$}}
    \raise1pt\hbox{$>$}}}                

\begin{document}
\includegraphics[width=3.5cm]{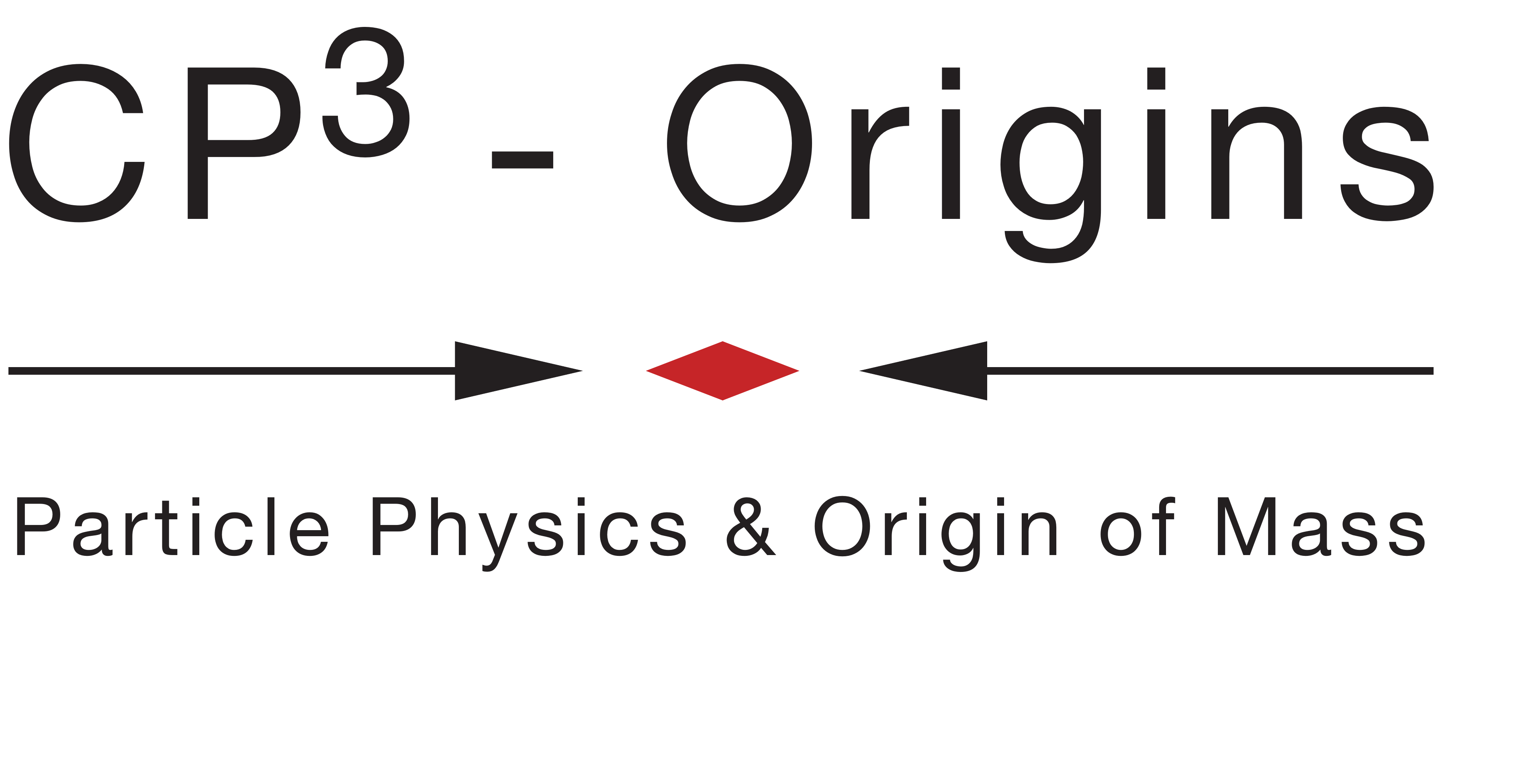}
\title{\Large  \color{rossoCP3} Minimal Composite Inflation }
\author{Phongpichit Channuie$^{\color{rossoCP3}{\varheartsuit}}$}\email{channuie@cp3-origins.net} 
\author{Jakob Jark J\o rgensen$^{\color{rossoCP3}{\varheartsuit}}$}\email{joergensen@cp3-origins.net} 
\author{Francesco Sannino$^{\color{rossoCP3}{\varheartsuit}}$}\email{sannino@cp3-origins.net} 
\affiliation{
$^{\color{rossoCP3}{\varheartsuit}}${ CP}$^{ \bf 3}${-Origins}, 
University of Southern Denmark, Campusvej 55, DK-5230 Odense M, Denmark.
} 
\begin{abstract}
 We investigate models in which the inflaton emerges as a composite field of a four dimensional, strongly interacting and nonsupersymmetric gauge theory featuring purely fermionic matter. We show that it is possible to obtain successful inflation via non-minimal coupling to gravity, and that the underlying dynamics is preferred to be near conformal. We discover that the compositeness scale of inflation is of the order of the grand unified energy scale.  \\ 
[.1cm]
{\footnotesize  \it Preprint: CP$^3$-Origins-2011-06}
 \end{abstract}

\maketitle
 
\section {Introduction}

 The standard model (SM) Higgs sector is plagued by the naturality problem, meaning that quantum corrections generate unprotected quadratic divergences requiring a huge fine-tuning if the model were to be true till the Planck scale. Such a fine tuning goes under the name of the hierarchy problem of the SM. Similarly the inflaton, the field needed to initiate a period of rapid expansion of our Universe, suffers from the same kind of untamed quantum corrections. It is therefore reasonable to ask if one can provide similar solutions to these two problems. We will show that it is possible to construct models in which the inflaton emerges as a composite state of a four-dimensional strongly coupled theory. 
 
As templates we use  models of dynamical electroweak symmetry breaking, also known as Technicolor, and reviewed in \cite{Sannino:2009za,Sannino:2008ha}. Here the Higgs sector of the SM is replaced by a new underlying four-dimensional gauge dynamics free from scalars. The Higgs is therefore composite and identified with a techni-hadron. The simplest models of Technicolor passing precision tests are known as Minimal Walking Technicolor models (MWT) \cite{Sannino:2004qp,Hong:2004td,Dietrich:2005wk,Dietrich:2005jn,Ryttov:2008xe,Frandsen:2009fs,Frandsen:2009mi,Foadi:2007ue,Belyaev:2008yj,Antola:2009wq}. We introduce a novel  gauge dynamics generating the inflaton field dynamically.  The inflaton potential is then identified with the low energy effective theory encoding the interactions for the lightest composite scalar. To be concrete we consider an underlying $SU(N)$ gauge theory featuring fermions transforming according to the adjoint representation of $SU(N)$. We consider two Dirac flavors corresponding to an $SU(4)$  global quantum symmetry of the theory. The reason behind this choice is fourfold: a] These kind of theories have a very interesting and rich dynamics \cite{Sannino:2004qp,Dietrich:2006cm,Ryttov:2007sr,Ryttov:2007cx,Ryttov:2008xe,Mojaza:2010cm,Pica:2010mt,Pica:2010xq,Mojaza:2011rw} being able with a small number of flavors to develop large distance conformality; b] Gauge theories with fermions transforming according to complex representations of the underlying gauge group possess a quantum global symmetry structure already contained in $SU(2N_f)$, and therefore we can recover their associated low energy effective theories easily in our framework; c] When investigating the possibility that the inflaton is the composite Higgs itself then the model is automatically the MWT one;  d] These theories are being subject to intensive numerical investigations via lattice simulations \cite{Catterall:2007yx, DelDebbio:2008wb,Shamir:2008pb,Deuzeman:2008sc,DelDebbio:2008zf,Catterall:2008qk,DelDebbio:2008tv,DeGrand:2008kx,Hietanen:2008mr,Appelquist:2009ty,Hietanen:2009az,Deuzeman:2009mh,Hasenfratz:2009ea,DelDebbio:2009fd,Fodor:2009wk,Fodor:2009ar,DeGrand:2009hu,Catterall:2009sb,Bursa:2009we,Bilgici:2009kh,Kogut:2010cz,Hasenfratz:2010fi,DelDebbio:2010hu,DelDebbio:2010hx,Catterall:2010du}.
 
 The paper is organized as follows: In Section \ref{framework} we first introduce the underlying gauge theory we employ to drive inflation and then present its associated relevant low energy effective theory for the composite inflaton. In Section \ref{coupling} we couple the effective theory non-minimally to gravity and show that the composite inflation scenario is viable. The non-minimal coupling to gravity occurs via four-fermion interactions at the underlying gauge theory level. We show that if the composite dynamics is near conformal with large anomalous dimensions for the fermion mass operator then it is possible to decouple the origin of these four fermion interactions from the inflationary dynamics. Intriguingly the composite scale for a successful composite inflation is predicted to be of the order of the grand unified theories energy scale. In the last Section we briefly summarize our findings.
 
\section{ Minimal Composite \& Conformal Inflation}
  \label{framework}
  
 We start with the basic assumption that the inflaton is not an elementary degree of freedom but a composite state of a strongly interacting nonsupersymmetric four-dimensional gauge theory featuring only fermionic matter. We make the further simplification to consider a single species of fermions transforming according to a given representation with respect to the underlying {\it techni-inflation} gauge group. Gauge theories can exist in different phases depending on the number of flavors, colors, matter and gauge representation. The collection of the possible phases as function of the number of flavors and colors for a given matter representation can be summarized by a phase diagram \cite{Sannino:2004qp,Dietrich:2006cm,Ryttov:2007sr,Ryttov:2007cx,Sannino:2009aw,Mojaza:2010cm,Pica:2010mt,Pica:2010xq,Mojaza:2011rw}. For a given matter representations there is a {\it conformal window}  in the color-flavor phase diagram, where the underlying gauge theory develops an infrared attractive fixed point. Within the conformal window the gauge theory displays large distance conformality. Near the lower end of the conformal window the specific gauge theory has a number of interesting properties: i) Gauge singlet correlators have (near) power law behaviors; ii) The spectrum of the theory features a light composite scalar compared to the composite scale of the theory   $\Lambda_{MCI} \simeq 4\pi v$, with $v$ the scale of the fermion condensate;  iii) Close to the lower end of the conformal window the anomalous dimension of the fermion mass is expected to be large. We have already mentioned that this is a welcome feature in the introduction; iv) A small number of flavors is needed to develop an infrared attractive fixed point for theories with matter in higher dimensional representations of the gauge group \cite{Sannino:2004qp}. In particular, theories with adjoint fermionic matter with two Dirac flavors are expected to be (near) conformal at large distances.  
   
 \subsection{Underlying Minimal Conformal Gauge Theory for Inflation}
We consider as underlying gauge theory for techni-inflation the $SU(N)$ gauge group with $N_f=2$ Dirac massless fermions transforming according to the adjoint representation of $SU(N)$. This theory has a quantum global symmetry $SU(4)$ expected to break spontaneously to $SO(4)$ when the fermion condensate forms. The associated effective Lagrangian has been constructed explicitly in \cite{Foadi:2007ue}\footnote{An equally interesting possibility is the use of pseudoreal representations of the underlying gauge group for which the expected pattern of symmetry breaking is $SU(2N_f) \rightarrow Sp(2N_f)$ which has been investigated in \cite{Appelquist:1999dq,Duan:2000dy} however our main physical results are general.}.  For $N=2$ we recover the MWT model, however for the composite inflation purpose any $N$ can be considered. Since the fermions transform according to the adjoint representation the size of the conformal window does not depend sensitively on $N$. Moreover at large number of colors we are guaranteed that the inflaton has a narrow width and therefore decoupled from the rest of the strongly coupled states making its effective description robust. However, we will not limit our analysis only to the large $N$ limit. To discuss the symmetry properties of the theory it is
convenient to use the Weyl basis for the underlying fermions and arrange them in the following vector transforming according to the
fundamental representation of SU(4)
\beq Q= \begin{pmatrix}
U_L \\
D_L \\
-i\sigma^2 U_R^* \\
-i\sigma^2 D_R^*
\end{pmatrix} \ ,
\label{SU(4)multiplet} \eeq where $U_L$ and $D_L$ are the left
handed techniup and technidown respectively, and $U_R$ and $D_R$ are
the corresponding right handed particles. We are using a Technicolor friendly notation to allow for a straightforward identification of these states with the ones relevant at the electroweak scale. Assuming the standard
breaking to the maximal diagonal subgroup, the SU(4) symmetry
spontaneously breaks to $SO(4)$. Such a breaking is driven by the
following condensate \beq \langle Q_i^\alpha Q_j^\beta
\epsilon_{\alpha \beta} E^{ij} \rangle =-2\langle \overline{U}_R U_L
+ \overline{D}_R D_L\rangle \ , \label{conde}
 \eeq
where the indices $i,j=1,\ldots,4$ denote the components
of the tetraplet of $Q$, and the Greek indices indicate the ordinary
spin. The matrix $E$ is a $4\times 4$ matrix defined in terms
of the 2-dimensional unit matrix as
 \beq E=\left(
\begin{array}{cc}
0 & \mathbbm{1} \\
\mathbbm{1} & 0
\end{array}
\right) \ . \eeq

Here 
$\epsilon_{\alpha \beta}=-i\sigma_{\alpha\beta}^2$ and $\langle
 U_L^{\alpha} {{U_R}^{\ast}}^{\beta} \epsilon_{\alpha\beta} \rangle=
 -\langle  \overline{U}_R U_L
 \rangle$. A similar expression holds for the $D$ techniquark.
The above condensate is invariant under an $SO(4)$ symmetry. This leaves us with nine broken  generators with associated Goldstone bosons. The fundamental Lagrangian replacing the inflaton one is:
\begin{eqnarray}
\mathcal{L}_{\rm Inflation}&\rightarrow &  -\frac{1}{4}{\cal F}_{\mu\nu}^a {\cal F}^{a\mu\nu} + i\bar{Q}_L
\gamma^{\mu}D_{\mu}Q_L + i\bar{U}_R \gamma^{\mu}D_{\mu}U_R +
i\bar{D}_R \gamma^{\mu}D_{\mu}D_R  
 \end{eqnarray}
with the techni-inflation field strength ${\cal F}_{\mu\nu}^a =
\partial_{\mu}{\cal A}_{\nu}^a - \partial_{\nu}{\cal A}_{\mu}^a + g_{TC} \epsilon^{abc} {\cal A}_{\mu}^b
{\cal A}_{\nu}^c,\ a,b,c=1,\ldots,N^2 - 1$.
For the left handed technifermions the covariant derivative might or might not include the SM fields. This model becomes MWT if $N=2$.  We gauge the left and right symmetries appropriately and further identify the techni-inflation with the composite Higgs. The MWT covariant derivative reads: 
\begin{eqnarray}
D_{\mu} Q^a_L &=& \left(\delta^{ac}\partial_{\mu} + g_{TC}{\cal
A}_{\mu}^b \epsilon^{abc} - i\frac{g}{2} \vec{W}_{\mu}\cdot
\vec{\tau}\delta^{ac} -i g'\frac{y}{2} B_{\mu} \delta^{ac}\right)
Q_L^c \ .
\end{eqnarray}
${\cal A}_{\mu}$ are the techni gauge bosons, $W_{\mu}$ are the
gauge bosons associated to SU(2)$_L$ and $B_{\mu}$ is the gauge
boson associated to the hypercharge. $\tau^a$ are the Pauli matrices
and $\epsilon^{abc}$ is the fully antisymmetric symbol. In the case
of right handed techniquarks the third term containing the weak
interactions disappears and the hypercharge $y/2$ has to be opportunely modified 
according to whether it is an up or down techniquark to avoid gauge anomalies.  

\subsection{Scalar Sector of Minimal Conformal Inflation (MCI) }\label{sec:scalar}

For  models of composite inflation, as it is for the case of models of dynamical breaking of the electroweak symmetry, it is convenient to introduce a low energy effective theory for the relevant degrees of freedom.  We borrow, for the composite inflation effective Lagrangian, the one constructed in \cite{Foadi:2007ue} for MWT and rename it the Minimal Conformal Inflation (MCI) effective theory. We have already mentioned earlier that we can use larger number of colors with respect to models of dynamical electroweak symmetry breaking. For the latter electroweak precision data limit the size of the underlying gauge group. A large number of colors guarantees the inflaton to be a narrow state with a decay width vanishing as $1/N^2$. 

The relevant effective theory for composite inflation consists, in our model, of a composite inflaton and its pseudoscalar partner, as well as nine pseudoscalar Goldstone bosons and their scalar partners. These
can be assembled in the matrix
\begin{eqnarray}
M = \left[\frac{\sigma+i{\Theta}}{2} + \sqrt{2}(i\Pi^a+\widetilde{\Pi}^a)\,X^a\right]E \ ,
\label{M}
\end{eqnarray}
which transforms under the full SU(4) group according to
\begin{eqnarray}
M\rightarrow uMu^T \ , \qquad {\rm with} \qquad u\in {\rm SU(4)} \ .
\end{eqnarray}
The $X^a$'s, $a=1,\ldots,9$ are the generators of the SU(4) group which do not leave  the vacuum expectation value (VEV) of $M$ invariant
\begin{eqnarray}
\langle M \rangle = \frac{v}{2}E
 \ .
\end{eqnarray}
Note that the notation used is such that $\sigma$ is a \emph{scalar}
while the $\Pi^a$'s are \emph{pseudoscalars}. It is convenient to
separate the fifteen generators of SU(4) into the six that leave the
vacuum invariant, $S^a$, and the remaining nine that do not, $X^a$.
Then the $S^a$ generators of the SO(4) subgroup satisfy the relation
\begin{eqnarray}
S^a\,E + E\,{S^a}^{T} = 0 \ ,\qquad {\rm with}\qquad  a=1,\ldots  ,  6 \ ,
\end{eqnarray}
so that $uEu^T=E$, for $u\in$ SO(4). The explicit realization of the generators is shown in the appendix of \cite{Foadi:2007ue}.
With the tilde fields included, the matrix $M$ is invariant in form under U(4)$\equiv$SU(4)$\times$U(1)$_{\rm
A}$, rather than just SU(4). However the U(1)$_{\rm A}$ axial symmetry is anomalous, and is therefore broken at the quantum level.

The connection between the composite scalars and the underlying technifermions can be derived from the transformation properties under SU(4), by observing that the elements of the matrix $M$ transform like technifermion bilinears:
\begin{eqnarray}
M_{ij} \sim Q_i^\alpha Q_j^\beta \varepsilon_{\alpha\beta} \quad\quad\quad {\rm with}\ i,j=1\dots 4.
\label{M-composite}
\end{eqnarray}
The MCI Lagrangian is
\begin{eqnarray}
{\cal L}_{\rm MWT} &=& \frac{1}{2}{\rm Tr}\left[D_{\mu}M D^{\mu}M^{\dagger}\right] - {\cal V}(M) \ ,
\label{Letc}
\end{eqnarray}
where the potential reads
\begin{eqnarray}
{\cal V}(M) & = & - \frac{m^2}{2}{\rm Tr}[MM^{\dagger}] +\frac{\lambda}{4} {\rm Tr}\left[MM^{\dagger} \right]^2 
+ \lambda^\prime {\rm Tr}\left[M M^{\dagger} M M^{\dagger}\right] \nonumber \\
& - & 2\lambda^{\prime\prime} \left[{\rm Det}(M) + {\rm Det}(M^\dagger)\right] \ ,
\label{Vdef}
\end{eqnarray}
The potential ${\cal V}(M)$ is SU(4) invariant. It produces a VEV
which parameterizes the techniquark condensate, and spontaneously
breaks SU(4) to SO(4). In terms of the model parameters the VEV is
\begin{eqnarray}
v^2=\langle \sigma \rangle^2 = \frac{m^2}{\lambda + \lambda^\prime - \lambda^{\prime\prime} } \ ,
\label{VEV}
\end{eqnarray}
while the inflaton mass is:
\begin{eqnarray}
M_I^2 = 2\ m^2 \ .
\end{eqnarray}
The linear combination $\lambda + \lambda^{\prime} -
\lambda^{\prime\prime}$ corresponds to the composite inflaton self coupling. 
We have nine Goldstones which migh or not acquire any mass or, be absorbed by gauging some of the global symmetries of the theory as it happens for some of the Goldstones when the MCI is identified with the MWT model.  The remaining scalar and pseudoscalar masses are
\begin{eqnarray}
M_{\Theta}^2 & = & 4 v^2 \lambda^{\prime\prime} \nonumber \\
M_{A^\pm}^2 = M_{A^0}^2 & = & 2 v^2 \left(\lambda^{\prime}+\lambda^{\prime\prime}\right) \nonumber \\
M_{\widetilde{\Pi}_{UU}}^2 & = & M_{\widetilde{\Pi}_{UD}}^2 = M_{\widetilde{\Pi}_{DD}}^2 =
  2 v^2 \left(\lambda^{\prime} + \lambda^{\prime\prime }\right) \ .
\end{eqnarray}
 To gain further insight in some of the mass relations one can use different types of  large N limits investigated in \cite{Sannino:2009za,Hong:2004td}. Besides the techni-scalar sector we expect other higher spin bound states to appear in the low energy effective theory.   We have already shown how to include these states at the effective Lagrangian level in \cite{Foadi:2007ue}. We focus here on the scalar sector since it is the most relevant for the inflationary paradigm but intend to investigate the effects of the spin one states in the future. 
An interesting feature of these models is the presence of a potentially light composite inflaton with respect to the scale dynamical  $4\pi v$. This point has been stressed in \cite{Hong:2004td,Sannino:2008ha} using Large N arguments and supersymmetry, and in \cite{Dietrich:2005jn,Dietrich:2006cm}, using the saturation of the trace of the energy momentum tensor. More generally it was argued that  models  featuring (near) conformal dynamics  contain a composite scalar state which is light with respect to the new strongly coupled scale ($4\,\pi \, v$).     Recent investigations using Dyson-Schwinger (SD) \cite{Doff:2008xx} and gauge-gravity dualities \cite{Fabbrichesi:2008ga} also arrived to the conclusion that these composite state is light.
 
\section{Minimal Conformal Inflaton  Non-Minimally Coupled to Gravity}
\label{coupling}

It was proposed in \cite{Bezrukov:2007ep} that the inflationary expansion of the early Universe can be linked to the SM by identifying the SM Higgs boson with the inflaton. The salient feature of the Higgs-inflation mechanism is the non-minimal coupling of the Higgs  doublet field ($H$) to gravity. This happens by adding a term of the type $\xi H^{\dagger} H R$ to the standard gravity-matter action, with $\xi$ a new coupling constant. This non-minimal coupling of scalar fields to gravity has a long history \cite{Spokoiny:1984bd,Futamase:1987ua,Salopek:1988qh,Fakir:1990eg,Kaiser:1994vs,Komatsu:1999mt,Tsujikawa:2004my}. A nonzero value of $\xi$ is needed since for $\xi=0$ an unacceptably large amplitude of primordial inhomogeneities is generated for a realistic quartic Higgs self-interaction term \cite{Linde:1983gd}. It was found in \cite{Bezrukov:2007ep} that with $\xi$ of the order $10^4$ the model leads to successful inflation, provides the graceful exit from it, and produces the spectrum of primordial fluctuations in good agreement with the observational data. This scenario was further explored in \cite{Barvinsky:2008ia,Bezrukov:2008ut,GarciaBellido:2008ab,DeSimone:2008ei,Bezrukov:2008ej,Bezrukov:2009db,Barvinsky:2009fy}. It was, however, noted  in \cite{Burgess:2009ea,Barbon:2009ya,Burgess:2010zq} that the operator describing non-minimal coupling, when written via canonically normalized fields, has dimension five suppressed by the scale $\Lambda_0 = M_P / \xi$ with $M_P$ the planck scale. If $\Lambda_0$ were to be identified with the ultra-violet (UV) cutoff, above which the SM has to be replaced by a more fundamental theory the Higgs inflation scenario would be technically {\it unnatural} since for large $\xi$ the scale $\Lambda_0$ is considerably lower than the Planck mass. Moreover the value of the Hubble expansion rate during inflation is close to $\Lambda_0$ making the contribution of the unknown effects coming from the physics beyond the SM sizable \cite{Burgess:2009ea}. This potential unnaturalness has been carefully reanalyzed in \cite{Bezrukov:2010jz}. Here it was stressed that this framework can be made consistent by viewing this approach as a description with a cutoff scale depending on the energy scale where the effective theory is active \cite{Cheung:2007st,Weinberg:2008hq}. Here we would like to use this framework in order to test the hypothesis that a  composite model of inflation can serve as a natural model for a rapid expansion of the Universe. The Higgs Lagrangian is now identified with the MCI effective theory which we couple non-minimally to gravity in the Jordan frame as follow:
\begin{align}
\mathcal{S}_{\text{J,MCI}}=\int d^{4}x \sqrt{-g}\left[- \frac{M^{2}_{P}}{2}R - \frac{1}{2}\xi \text{Tr}\left[\text{M}\text{M}^{\dagger}\right]R+\mathcal{L}_{\text{MCI}}\right]. \label{tcsj}
\end{align}
The non-minimally coupled term in the Lagrangian corresponds at the fundamental level to a four-fermion interaction term coupled to the Ricci scalar in the following way: 
\begin{equation}
\frac{\xi}{2} \frac{ (	Q Q)^{\dagger} Q Q}{\Lambda_{ECI}^4} \, R \ ,
\end{equation}
with $\Lambda_{ECI} \geq 4\pi v   $ a new high energy scale where this operator generates via some yet to be specified new dynamics termed here Extended Conformal Inflation (ECI) dynamics whose details are not relevant for the present discussion. Using the renormalization group equation for the chiral condensate we have:
\begin{equation}
\langle QQ\rangle_{\Lambda_{ECI}} \sim \left(\frac{\Lambda_{ECI}}{\Lambda_{MCI} } \right)^{\gamma} \langle QQ \rangle_{\Lambda_{MCI} } \ , 
  \end{equation}
 where the subscript indicates the energy at which the operators are evaluated and $\Lambda_{MCI} = 4\pi v$. We assumed to underlying theory to be near conformal in the energy range $\Lambda_{MCI}  \leq \mu \leq \Lambda_{ECI}$  and therefore $\gamma$  is almost constant.  If the fixed value of $\gamma$ is around two  the explicit dependence on the $\Lambda_{ECI}$ disappears since $M \sim \langle QQ \rangle_{\Lambda_{MCI}} /\Lambda_{MCI} ^2$.  In other words for $\gamma$ around two the ECI dynamics decouples from the lower energy inflationary physics.
 
The inflaton is identified with the field $\sigma$ and we drop the other fields in $M$. The composite inflaton effective action reads: 
\begin{align}
\mathcal{S}_{\text{J,MCI}}=\int d^{4}x \sqrt{-g}\left[-\frac{M^{2}_{P}}{2}\Omega^{2}R + \frac{1}{2}g^{\mu\nu}\partial_{\mu}\sigma\partial_{\nu}\sigma+\frac{m^{2}}{2}\sigma^{2}-\frac{\kappa}{4}\sigma^{4}\right], \label{sjtc}
\end{align}
where
\begin{align}
\Omega^{2}=\left(\frac{M^{2}_{P}+\xi\sigma^{2}}{M^{2}_{P}}\right) \ , \quad {\rm and } \quad  
\kappa=\left(\lambda+\lambda^{'}-\lambda^{''}\right). \label{kap} 
\end{align}
By applying the conformal transformation $g_{\mu \nu} \rightarrow \tilde{g}_{\mu \nu} = \Omega ^2 g_{\mu \nu}$ we eliminate the non-minimal coupling between $\sigma$ and the gravitational field. The resulting action in the Einstein frame is:
 \begin{align}
S_{\text{E,MCI}}=\int d^{4}x \sqrt{-\tilde{g}}\left[- \frac{M^{2}_{P}}{2}\tilde{R}+\frac{1}{2}\left(\frac{M^{2}_{P}\Omega^{2}+6\xi^{2}\sigma^{2}}{M^{2}_{P}\Omega^{4}}\right)\tilde{g}^{\mu\nu}\partial_{\mu}\sigma\partial_{\nu}\sigma+\Omega^{-4}\left(  \frac{m^2}{2}\sigma^2 - \frac{\kappa}{4}\sigma^{4}\right)\right] \ . \label{action}
\end{align}
Having removed the non-minimal coupling to gravity we landed with a non-canonical kinetic term for the scalar field $\sigma$ which can be put in a canonical form by introducing the following field  $\chi(\sigma)$ linked to $\sigma$ via:
\begin{align}
\frac{1}{2}\tilde{g}^{\mu\nu}\partial_{\mu}\chi(\sigma)\partial_{\nu}\chi(\sigma)=\frac{1}{2}\left(\frac{d\chi}{d\sigma}\right)^{2}\tilde{g}^{\mu\nu}\partial_{\mu}\sigma\partial_{\nu}\sigma\ , \label{chi}
\end{align}
where
\begin{align}
\chi'= \frac{d\chi}{d\sigma}=\sqrt{\frac{M^{2}_{P}\Omega^{2}+6\xi^{2}\sigma^{2}}{M^{2}_{P}\Omega^{4}}}\ ,\label{difchi}
\end{align}
and the action now reads:  
\begin{align}
S_{\text{E,MCI}}=\int d^{4}x \sqrt{-\tilde{g}}\left[-\frac{1}{2}M^{2}_{P}\tilde{R}+\frac{1}{2}\tilde{g}^{\mu\nu}\partial_{\mu}\chi(\sigma)\partial_{\nu}\chi(\sigma)+\Omega^{-4}\left(  \frac{m^2}{2}\sigma^2 - \frac{\kappa}{4}\sigma^{4}\right)\right] \ . \label{newaction}
\end{align}
 
 We consider  the large field regime:
\begin{align}
\sigma>>\frac{M_{p}}{\sqrt{\xi}} \ . \label{largesigma}
\end{align}
for which the following approximations hold
\begin{align}
\Omega^{2} \approx\frac{\xi\sigma^{2}}{M^{2}_{p}}, \qquad  \qquad   \frac{d\chi}{d\sigma}     \approx\sqrt{6}\frac{M_{p}}{\sigma}  \ . \label{largeome}
\end{align}
The solution to the second equation to the right reads:  
\begin{align}
\chi =\sqrt{6}M_{P}\ln(\frac{\sqrt{\xi}\sigma}{M_{P}}) \ . \label{newchi}
\end{align}
The potential becomes: \begin{align}
U(\chi)=\frac{\kappa}{4}\frac{M^{4}_{P}}{\xi^{2}}\left(1+e^{\frac{-2\chi}{\sqrt{6}M_{P}}}\right)^{-2}  = \frac{\kappa}{4}\frac{M^{4}_{P}}{\xi^{2}}\left(1+\frac{M^{2}_{P}}{\xi\sigma^{2}}\right)^{-2} \ . \label{largechipot}
\end{align}
The associated slow-roll parameters are: 
 \begin{align}
\epsilon =\frac{M^{2}_{P}}{2}\left(\frac{dU/d\chi}{U}\right)^{2} = \frac{M_p ^2}{2} \left( \frac{U'}{U} \frac{1}{\chi'} \right) ^2 , \label{epsil}
\end{align}
and
\begin{align}
\eta =M^{2}_{P}\left(\frac{d^{2}U/d\chi^{2}}{U}\right) = M_p ^2 \frac{U '' \chi'  - U' \chi'' }{U {\chi'}^3} \ .\label{eta}
\end{align}
Here $U'$ denotes derivative with respect to $\sigma$.  We obtain
\begin{align}
\epsilon \simeq \frac{4}{3}e^{\left(-\frac{4\chi}{\sqrt{6}M_{p}}\right)} \ , \qquad \qquad 
\eta \simeq \frac{4}{3}e^{\left(-\frac{2\chi}{\sqrt{6}M_{p}}\right)}\ .  
\label{newepsil}
\end{align}
 In terms of the $\sigma$ field:
\begin{align}
\epsilon \simeq \frac{4}{3}\frac{M^{4}_{P}}{\xi^{2}\sigma^{4}}\ , \qquad \qquad \eta \simeq \frac{4}{3}\frac{M^{2}_{P}}{\xi\sigma^{2}}\ . 
 \label{epsilsig}
\end{align}
Inflation ends when $\epsilon=1$, i.e.
\begin{align}
\epsilon=1\simeq\frac{4}{3}\frac{M^{4}_{P}}{\xi^{2}\sigma_{end}^{4}} \quad \Rightarrow  \quad \sigma_{end}\simeq (\frac{4}{3})^{1/4}\frac{M_{P}}{\sqrt{\xi}}\simeq 1.07 \,  \frac{M_{P}}{\sqrt{\xi}}\ . \label{epsilend}
\end{align}

The number of e-foldings during inflation is:
\begin{align}
{\cal N}=\frac{1}{M^{2}_{P}}\int^{\chi_{ini}}_{\chi_{end}}\frac{U}{dU/d\chi}d\chi = \frac{1}{M^{2}_{P}}\int^{\sigma_{ini}}_{\sigma_{end}}\frac{U}{dU/d\sigma}\left(\frac{d\chi}{d\sigma}\right)^{2}d\sigma \simeq \frac{6}{\left(8M^{2}_{P}/\xi\right)}\left(\sigma^{2}_{ini}-\sigma^{2}_{end}\right) \ ,\label{enum1}
\end{align}
which combined with \eqref{epsilend} allows us to write: 
\begin{align}
\sigma_{ini}=\sqrt{\left(\frac{8{\cal N}}{6}+(1.07)^{2}\right)}\,\,\frac{M_{P}}{\sqrt{\xi}}\ . \label{inisig}
\end{align}
By requiring ${\cal N} \sim 60$ we get
\begin{align}
\sigma_{ini} \sim  9 \,\frac{M_{P}}{\sqrt{\xi}} \ . \label{appinisig}
\end{align} 

To generate the proper amplitude of the density perturbations the potential must satisfy at $
\sigma_{\rm WMAP}$ the normalization condition \cite{Bezrukov:2008ut}: 
\begin{equation}
\frac{U}{\epsilon} \simeq (0.0276 \, M_p)^4 \ , 
\end{equation}
corresponding to the initial value  assumed by the inflaton. We therefore deduce: 
\begin{equation}
\xi = \frac{{\cal N}}{(0.0276)^2}\sqrt{\frac{\kappa}{3}}  \sim 46000 \sqrt{\kappa}\ .
\end{equation}
For a strongly coupled theory we expect $\kappa$ to be of the order of unity and therefore $\xi \sim 46000$. This analysis resembles very closely the one for the SM Higgs inflation, except that our effective theory for the composite inflaton cannot be utilized for arbitrary large value of scalar field. The effective theory  is valid for: 
\begin{equation}
 \sigma < 4\pi v \  ,
\end{equation}
implying\begin{equation}
v >  \frac{9\,M_P}{4\pi \sqrt{\xi}} \sim (0.81 - 4.07) \times 10^{16} ~{\rm  GeV}\ .
\end{equation}
with the lower value obtained for the reduced Planck mass of $2.44\times 10^{18}$~GeV and the higher one for the standard one of $1.22\times 10^{19}$~GeV. This value is surprisingly close to the typical grand unification scale $M_{GUT}$ of $10^{16}$~GeV compatible with the phenomenologically viable proton decay time. This phenomenological constraint on $v$ forbids the identification of the composite inflaton with  the composite Higgs. This lower bound on the scale of composite inflation arises  from having assumed the effective theory to be valid during the inflationary period. This bound may be weakened if we consider directly the underlying strongly coupled gauge theory, however, this is beyond the scope of this initial investigation. First principle lattice computations could explore directly this possibility. Models of dynamical electroweak symmetry breaking, such as MWT, also lead to unification scenarios with a similar grand unified energy scale \cite{Gudnason:2006mk}.

\section{Conclusions}
\label{conclusions}

 We proposed models in which the inflaton is as a composite field stemming from a four dimensional strongly interacting nonsupersymmetric gauge theory. We demonstrated that it is possible to obtain a successful inflation. Quite surprisingly we discovered the composite scale to be the one typically associated to grand unified theories.  
  Because the scale of inflation is the grand unified one the composite inflaton cannot be identified with the composite Higgs state emerging in models of dynamical electroweak symmetry breaking.  It would be interesting to explore in the future possible links to models of holographic composite inflation \cite{Evans:2010tf}. Our results lead to the fundamental conclusion that different strongly coupled nonsupersymmetric gauge theories featuring fermionic matter can naturally account for dynamical breaking of the electroweak symmetry, dark matter and a dynamical origin of inflation.

\acknowledgments
We thank Simon Catterall and Claudio Pica for reading the manuscript and useful discussions. P. Channuie thanks the Royal Thai Government under the program {\it Strategic Scholarships for Frontier Research Network} of Thailand's Commission on Higher Education.

\end{document}